\documentclass[pre,amssymb,amsmath,amsfonts,superscriptaddress,twocolumn,showpacs]{revtex4}
\usepackage{graphicx,color}
\usepackage{color} 
\usepackage{hyperref}

\begin{document}

\title{Collective Motion with Anticipation:  Flocking, Spinning, and Swarming}
\author{Alexandre Morin}
\affiliation{Laboratoire de Physique de l'Ecole Normale Sup\'erieure de Lyon, Universit\'e de Lyon, CNRS, 46, all\'ee d'Italie, 69007 Lyon, France}
\author{Jean-Baptiste Caussin}
\affiliation{Laboratoire de Physique de l'Ecole Normale Sup\'erieure de Lyon, Universit\'e de Lyon, CNRS, 46, all\'ee d'Italie, 69007 Lyon, France}
\author{Christophe Eloy}
\affiliation{Aix Marseille Universit\'e, CNRS, Centrale Marseille, IRPHE UMR 7342, 13384, Marseille, France}
\author{Denis Bartolo}
\affiliation{Laboratoire de Physique de l'Ecole Normale Sup\'erieure de Lyon, Universit\'e de Lyon, CNRS, 46, all\'ee d'Italie, 69007 Lyon, France}

\begin{abstract}
We investigate the collective 
{\color{black} dynamics} of self-propelled particles able to probe and anticipate the orientation of their neighbors. We show that  a simple anticipation strategy hinders the emergence of {\color{black} homogeneous flocking patterns}. Yet, anticipation promotes two other forms of self-organization: collective spinning and swarming. In the  {spinning phase, all  particles} follow synchronous circular orbits, while in the { swarming phase,} the population condensates into a single compact swarm that cruises coherently without requiring any cohesive interactions. We quantitatively characterize and rationalize these phases of polar active matter and  {discuss}  potential applications {to the design of swarming robots}.
\end{abstract}

\maketitle
\section{introduction}
Over the last 20 years physicists {have devoted}  significant efforts to elucidate  the numerous dynamical patterns observed in populations of living organisms. Flocking patterns are a prominent example.  They refer to the self-organization of an ensemble of motile individuals into a homogeneous group undergoing coherent directed motion. As initially suggested by Vicsek et al.~\cite{Vicsek}, simple alignment interactions are sufficient to trigger the emergence of flocking patterns. Alinement-induced flocks have been demonstrated  theoretically, numerically and in  synthetic experiments~\cite{Vicsek_review,Marchetti_review,Bricard2013,Deseigne2010,Bausch2010,Ramaswamy2014}. 
In addition, quantitative analysis of animal trajectories ~\cite{Cavagna_review,Couzinscience,Theraulazfishes,Hemelrijk2012} support that the large scale features of some animal flocks can be rationalized by simple behavioral rules at the individual level, including {\color{black} velocity-alignment interactions.}
However, inferring a model from actual data requires a minimal set of hypothesis on the possible form of the interactions~\cite{Couzin_Review}, or on the statistical properties of the observables~\cite{Cavagna_review}. Until now, the overwhelming majority of the available models has been merely restrained to couplings between the {\em instantaneous} positions and orientations of neighboring individuals (see~\cite{Vicsekacceleration,Vicsekinertia,Theraulazfishes} for noteworthy exceptions). Investigating alternative dynamical rules may provide further insight into the collective dynamics of motile individuals.

Here, we generalize the conventional description of polar active matter~\cite{Marchetti_review}. We consider the dynamics of motile particles, {\color{black} which aligns their velocity with the orientation of their neighbors by anticipating their rotational motion}.
Naively,  {\color{black}anticipation would be expected to yield more robust flocks}. In striking contrast, we show that  simple anticipation rules result in much richer collective behaviors: 
{\color{black}Firstly,  when motile particles strongly anticipate the orientational changes, the population simultaneously breaks a continuous and a discrete symmetry.} The system  self-organizes into a spinning state where all the particles follow  closed circular orbits in a synchronized fashion. 
{\color{black}Secondly, a combination of alignment and moderate anticipation results in the emergence of a polar-liquid phase akin to the  flocking pattern observed without anticipation.}
{\color{black}Finally, at the onset of the flocking-to-spinning transition, the population forms stable compact swarms  despite the absence of any attractive couplings.}

{\color{black}The paper is organized as follows: we first introduce the equations of motion  of  the interacting self-propelled particles. The alignement and anticipation rules are described.  We then discuss the emergence of  the three ordered phases (spinning, flocking, and swarming phases). They} are quantitatively characterized, explained and compared to the conventional  patterns of polar active matter. 
We close this paper from an engineering perspective. We show that minimal anticipation rules can provide an effective and safe design strategy to {interrupt} the directed motion of a flock without shutting down the propulsion mechanism at the individual level. 

\section{Model: alignment and anticipation}
We consider an ensemble of $N$ self-propelled particles in {\color{black} a two-dimensional space}. The particle $i$, located at ${\bf r}_{i}(t)$, moves at a speed $v_0$ along the unit vector $\hat {\bf p}_i$ that makes an angle $\theta(\hat {\bf p}_i)\equiv\theta_i$ with the $x$-axis
\begin{align}
\dot {\bf r}_i(t)=v_0 \hat {\bf p}_i.
\label{eqr}
\end{align}
 {\color{black} The equation of motion of their instantaneous orientation defines their anticipation and alignement rules:} 
 \begin{align}
\dot \theta_i(t)=-\frac{1}{\tau}\big\langle\sin\left[{\theta_i-\left( \theta_j+\alpha \sigma_j\right )}\right]\big\rangle_{j \in\Omega_i}+\sqrt{2\eta}\xi_i(t),
\label{eqtheta}
\end{align}
where $\alpha$ is a scalar parameter and $\sigma_j\equiv\dot\theta_j/|\dot\theta_j|$ is the sign of the angular velocity.
The particles have a finite interaction radius $R$, and $\Omega_i$ {denotes} the ensemble of particles interacting with the $i^{\rm th}$ particle. 
 We henceforth refer to $\sigma_j$ as the spin of the particle $j$.
 This quantity is akin to the spin variable defined in~\cite{Vicsekinertia} up to a multiplicative factor, which is the local curvature of the particle trajectory. In Eq.~\ref{eqtheta}, the angular noises $\xi_i$'s are uncorrelated Gaussian random variables of unit variance, and $\tau$ is an orientational relaxation time. For the sake of {simplicity}, units are chosen so that $\tau=1$ and $R=1$, {and  the  control parameters left are $N$, $v_0$, $\alpha$, $\eta$, and the system size $L$.} 

Equations \ref{eqr} and \ref{eqtheta} have a simple physical meaning. When $\alpha=0$, they {reduce} to the continuous-time version of the seminal Vicsek model~\cite{peruani}: the particles interact via effective torques that promote alignment with the instantaneous orientation of the neighboring particles. Note that neither the momentum nor the angular momentum  is conserved in Eqs.~1 and~2.
 When {\color{black} $\alpha>0$,  the mean orientation of the neighbors}  is anticipated in a  simple fashion. If the  orientation of the $j^{\rm th}$ particle rotates in the  clockwise (resp. anti-clockwise) direction, {\color{black}an effective torque promotes the alignment along the direction} $\theta_j-\alpha$ (resp. $\theta_j+\alpha$). {\color{black}$\alpha$ is a constant angle used to anticipate the future orientation of the neighboring particles.}
 
In order to gain a better insight into this interaction scheme, we expand the sine functions and use elementary algebra to recast  Eq.~\ref{eqtheta} into
 \begin{align}
\dot \theta_i(t)=&-\cos\alpha\,\big\langle\sin\left({\theta_i-\theta_j}\right)\big\rangle_{j \in\Omega_i}
\label{vsalpha}
\\
&-\sin\alpha\,\big\langle\sin\left[{\theta_i- (\theta_j+\sigma_j\frac{\pi}{2})}\right]\big\rangle_{j \in\Omega_i}+\sqrt{2\eta}\xi_i(t),\nonumber
\end{align}
which {is the linear superposition of two models: a continuous-time Vicsek model and a $\alpha=\frac{\pi}{2}$ model}. 
The  second term on the r.h.s {of Eq.~\ref{vsalpha}} promotes  alignment with the acceleration of the surrounding particles. As the magnitude of the translational velocity is constant, the acceleration of a particle is perpendicular to its direction of motion: $\theta(\partial_t\hat {\bf p}_i)=\theta_i+\sigma_i\frac{\pi}{2}$.
Equation~\ref{vsalpha} is {\em qualitatively} similar to the discrete time model introduced in~\cite{Vicsekacceleration}.
We revisit here the impact of the velocity-acceleration coupling from a different perspective and further characterize the resulting phase {behavior}.

The special status of the  $\alpha=0$ and $\alpha=\frac{\pi}{2}$  models prompts us to characterize them separately. 
 However,
 a comprehensive numerical characterization would  require to vary systematically three independent parameters that are the propulsion speed $v_0$, the particle density and the noise magnitude $\eta$. In this paper we set $v_0=1/2$, which is a conventional value  that facilitates comparisons with previous polar active matter models~\cite{ChatePRL,ChatePRE}. If not mentioned otherwise, we provide results corresponding to a particle density {\color{black} $\rho=8$, corresponding to $N=2048$ particles}. Smaller density values yield qualitatively identical results, {but} require  longer convergence times for finite {values of $\alpha$}. We therefore focus solely on the impact of the anticipation {angle} $\alpha$ and  the angular noise $\eta$. 
 
 {\color{black}When $\alpha\neq0$ Eq. 2 and 3 are implicit as $\sigma_i$ depends on the time derivative of $\theta_i$. We solve Eqs.~\ref{eqr} and \ref{vsalpha} numerically using periodic boundary conditions in a square simulation box {of size $L$} and a forward Euler  method. This method  requires specifying the discretization scheme for the spin variables. Aiming at describing the impact of anticipation, we have naturally defined the discretized model by setting $\sigma_i(t)={\rm sign} \left[\theta_i(t)-\theta_i(t-\delta t)\right]$, where the time step $\delta t$ is  $10^{-2}\tau$ in all the simulations.}

 
 \section{Results and discussion}
 \subsection{{Case $\alpha=0$: Alignment-induced flocking.}}
 \begin{figure}
\begin{center}
\includegraphics[width=\columnwidth]{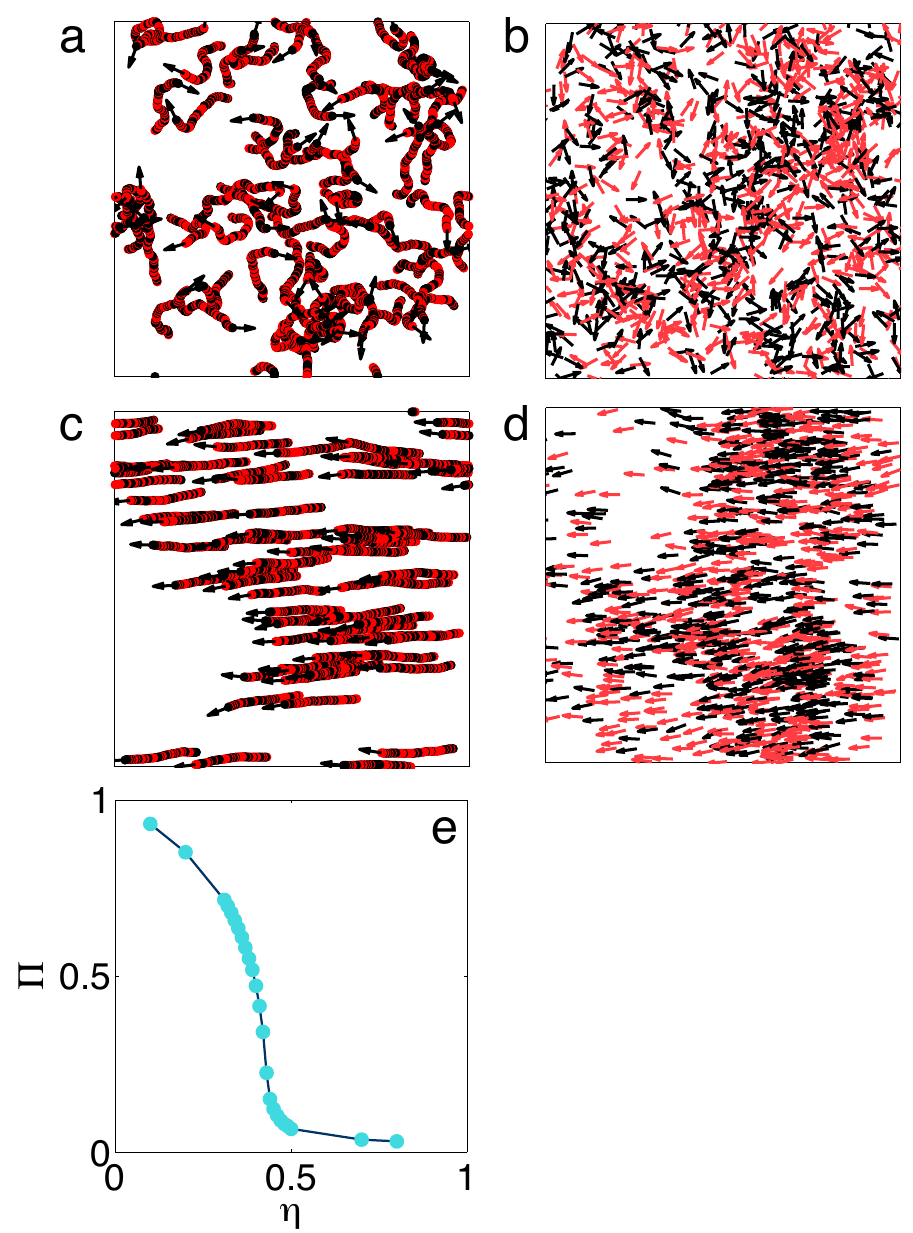}
\caption{Color online. Continuous-time Vicsek model ($\alpha=0$). (a) Superimposed plot of subsequent particle positions in the isotropic phase ($\eta=0.5$). The color accounts for the spin value. Red: $\sigma=-1$, and black: $\sigma=+1$.  (b)  Instantaneous orientations and positions of the particles in the isotropic phase ($\eta=0.5$). To improve clarity, only one half of all particles is shown. (c) and (d) Same plots for the flocking state ($\eta=10^{-2}$). (e) Polarization plotted as a function of the magnitude of the angular noise. For all the pictures and plots: $\rho=8$, $v_{0}=0.5$ and $L=16$ {\color{black}($N=2048$)}. }
\label{figvs}
\end{center}
\end{figure}
 {Here, we briefly recall  the phenomenology of the model when $\alpha=0$, which is equivalent to a continuous-time version of the classical Vicsek model~\cite{peruani}.}
 For high noise values,  the self-propelled particles undergo uncorrelated persistent random walks. The population forms an isotropic and homogeneous gas  depicted in Figs.~\ref{figvs}a, b. Decreasing the noise value below $\eta_{\rm 0}=0.45$, the rotational symmetry of the particle orientations is spontaneously broken: a macroscopic fraction of the population propel  in the same direction along straight trajectories, see Figs.~\ref{figvs}c,d, and  Supplementary Movie~1)~\cite{supp}. This flocking transition, is quantitatively captured by the sharp variation of the order parameter  $\Pi\equiv|\langle\hat{\bf p}_i\rangle_{i,t}|$ (the mean polarization), in Fig.~\ref{figvs}e. 
 
Note however that, unlike what is found in simulations of  discrete time Vicsek models (in the dilute limit)~\cite{ChatePRL,ChatePRE}, {we do not observe  the propagation of band-shape excitations  at the onset of collective motion.} The reason for this discrepancy is  purely  technical.  Based on previous numerical observations~\cite{ChatePRE} using an angular noise as in Eq.~\ref{vsalpha}, the emergence of localized bands  is expected to  occur only in extremely large systems. The relatively small size of our largest simulations ($L=16$, $N=2048$)  explains why  solitonic bands do not form here at the onset of collective motion.

The simulations in the case $\alpha=0$ reproduce the salient features of the Vicsek flocking transition, and  therefore validate our numerical scheme.  

 \subsection{{Case $\alpha=\frac{\pi}{2}$: Anticipation-induced spinning.}}
\begin{figure}
\begin{center}
\includegraphics[width=\columnwidth]{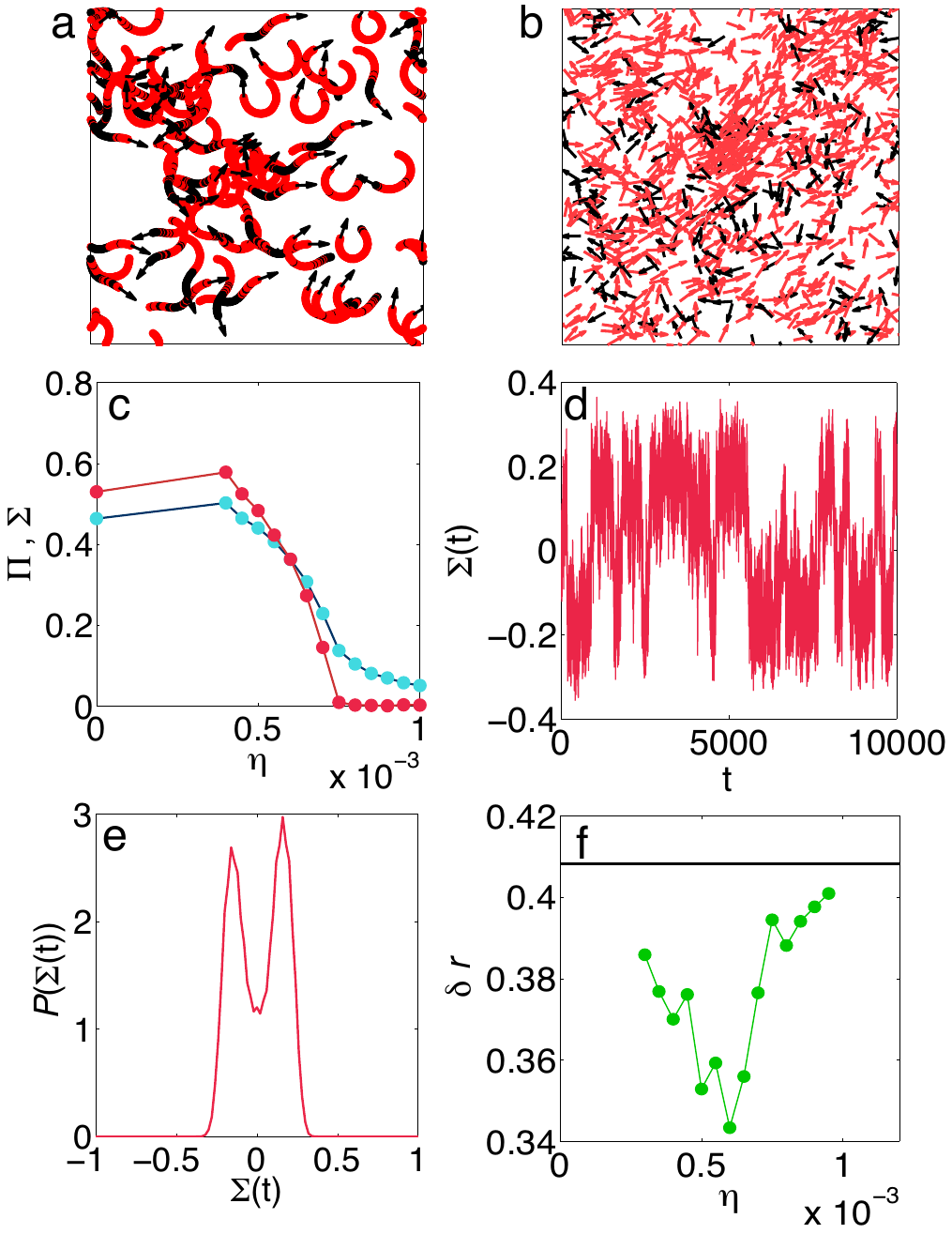}
\caption{Color online. {Case $\alpha=\frac{\pi}{2}$.} (a)Typical trajectories in the spinning phase  ($\eta=4\times 10^{-4}$). (b) Snapshot of the position and orientation of the motile particles for the same noise value. The color code accounts for the instantaneous spin as in Fig. 1 ($\eta=4\times 10^{-4}$). (c) Variations of the order parameters $\Pi$ (light/black symbols) and $\Sigma$ (dark/red symbols) with respect to the noise amplitude $\eta$.   (d) Time series of the instantaneous global spin $\Sigma(t)$, and (e) related probability distribution that displays a clear bimodal structure. (f) The spatial condensation at the transition is clearly seen from the variations of $\delta r$ with $\eta$, the horizontal line corresponds to the value of $\delta r$ for a uniform distribution of particles in a square box.  All the simulations correspond to $\rho=8$, $v_{0}=0.5$ and $L=16$ {\color{black}($N=2048$)}.}
\label{fig2pisur2}
\end{center}
\end{figure}

We {now} investigate the  anticipation model corresponding to $\alpha=\frac{\pi}{2}$. Not surprisingly, in the high noise regime, all the particles behave like uncorrelated persistent random walkers again, and form a homogeneous isotropic phase with no preferred spin,  Figs.~\ref{figvs}a and \ref{figvs}b. However, reducing the noise amplitude below $\eta_{\frac{\pi}{2}}=7\times 10^{-4}$, two symmetries are spontaneously broken: the continuous rotational symmetry of the particle orientation, together with the discrete $\mathbb Z_2$ symmetry of the spin variable. As {seen} see Figs.~\ref{fig2pisur2}a,b, a macroscopic fraction of the population synchronously follows persistent orbital trajectories with a constant radius, see also  Supplementary Movie~2~\cite{supp}. 

This alternative type of collective motion is well captured by two order parameters: the global polarization $\Pi$ and  the global spin $\Sigma$,  which is defined as the absolute value of the most probable instantaneous average spin $\langle\sigma_i(t)\rangle_i$. These two quantities sharply increase  as $\eta$ goes below $\eta_{\frac{\pi}{2}}$ as shown in Fig.~\ref{fig2pisur2}c. It is  also worth noting that, the temporal variations of  $\langle\sigma_i(t)\rangle_i$ is intermittent over time scales that are much larger than the individual relaxation time $\tau$ regardless of the noise amplitude, see Fig.~\ref{fig2pisur2}d. {\color{black} The rare intermittent events correspond to the  interruption of the  circular trajectories. The particles randomize their orientation at the same time and subsequently self-organize again to rotate synchronously, possibly along a different direction}. This long-time dynamics results in a bimodal structure of ${\cal P}(\langle\sigma_i(t)\rangle_i)$, seen in Fig.~\ref{fig2pisur2}e, and a posteriori justifies  defining the spin order parameter as the modulus of the most probable value of $\Sigma(t)$.

\begin{figure*}
\begin{center}
\includegraphics[width=\textwidth]{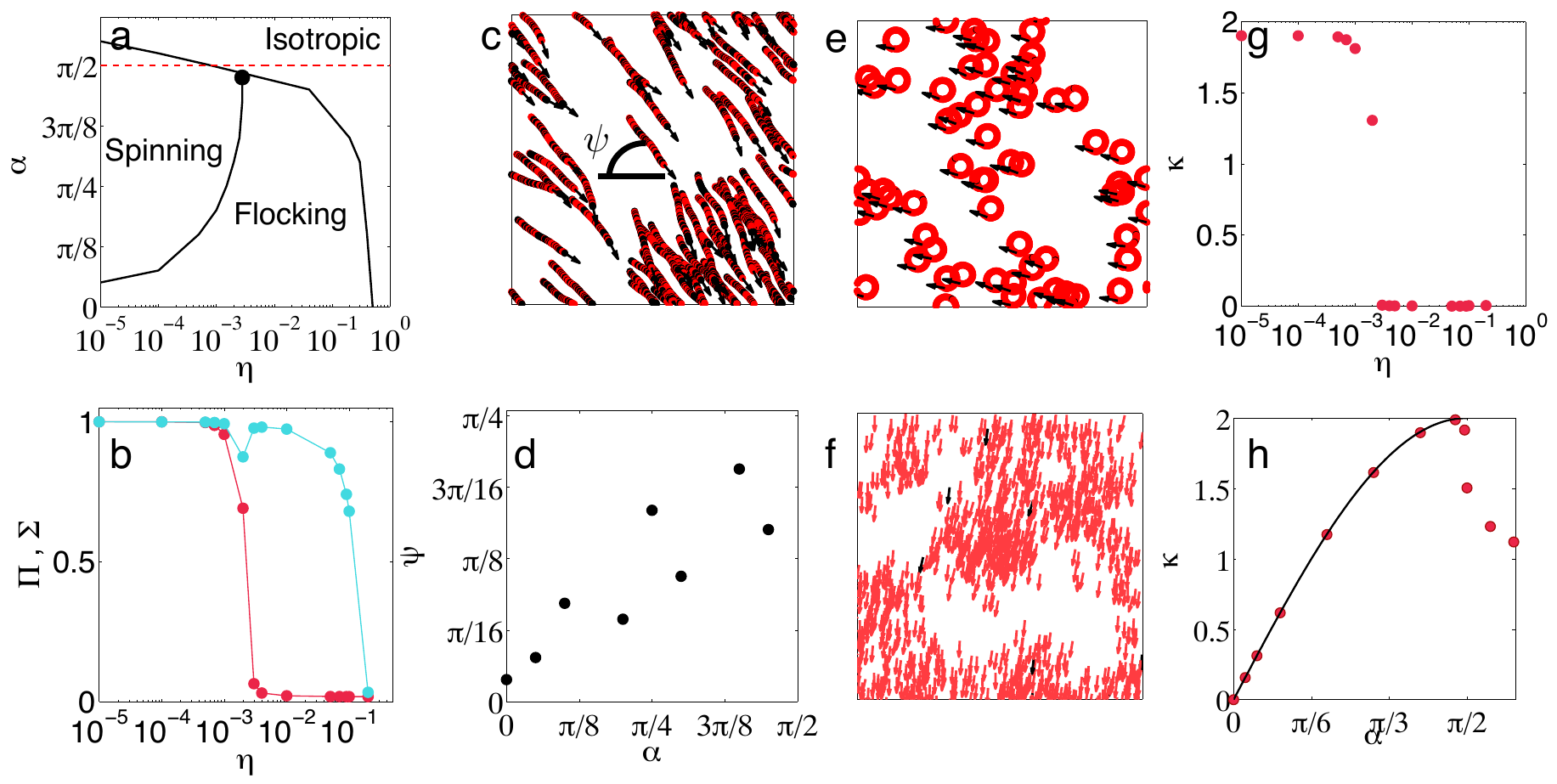}
\caption{Color online. Alignment and anticipation for finite $\alpha$. (a) Phase diagram showing the region where the isotropic, the flocking and the spinning phases are stable. (b) Sharp variations of the polarization and spin order parameters with $\eta$ {(same color code as in Fig.~\ref{fig2pisur2})}. (c) Exemples of particle trajectories for $\alpha=0.8\pi/2$ in the flocking state, for  $\eta = 10^{-2}$. (d) Variations of the angle $\psi$ between the polarization and the principal axis of the simulation box, as a function of the anticipation angle $\alpha$. (e) Typical spinning trajectories observed for $\eta= 7 \times 10^{-4}$. (f) Instantaneous position and orientation of one half of the particles, with the same color code for the spin variable. The entire population rotates in the same direction. (g) Variation of the curvature of the spinning trajectories with the noise amplitude for $\alpha=0.4\pi$. (h) Variation of the curvature of the spinning trajectories as a function of $\alpha$ in the zero-noise limit. Full line: analytical prediction $\kappa=\sin\alpha/v_0\tau$.  For all simulations, $\rho=8$, $v_{0}=0.5$, and $L=16$ {\color{black}($N=2048$)}.}
\label{fig3alpha1}
\end{center}
\end{figure*}

{In addition the individual circular trajectories observed in this spinning phase} are not robust: they continuously form and disrupt, and neither $\Pi$ nor $\Sigma$ converges to 1 as $\eta$ goes to 0 (Fig.~\ref{fig2pisur2}c). At any given time, a small yet finite fraction of particles does not follow synchronous circular trajectories even for vanishingly small noise.   
These unusual strong fluctuations in the weak-noise limit, together with the circular motion of the particle, will be better understood in the next section, when inspecting the  phase diagram of the  general $\alpha$ model.

Last,  we stress that at the onset of global spinning motion, the spatial distribution of the particles is heterogeneous. The population concentrates into a large denser region surrounded by a more dilute ensemble of spinning particles. This spontaneous breaking of the translational invariance is quantified by measuring the variations of $\delta r\equiv \left|\langle {\bf r}_i-\langle{\bf r}_i\rangle_i \rangle_{i,t}\right|$ as a function of the angular noise, Fig.~\ref{fig2pisur2}f . This spatial heterogeneity emerges in the absence of explicit couplings between the positional degrees of freedom, which is a very persistent feature of all the active-particle models involving any form of short-range interactions~\cite{Caussin2014}.  
 \subsection{{Case $\alpha>0$: From flocking to spinning and swarming.}}
 We now discuss the  general case, for which the anticipation angle $\alpha$ {\color{black} can take any positive value}. The collective behavior of the population is summarized by a phase diagram that divides the $(\eta,\alpha)$ plane into three regions as shown in Fig.~\ref{fig3alpha1}. The boundaries of the phase diagram correspond to the points where the order parameters, $\Pi$ or $\Sigma$, go from  zero to a finite value. 
 When $0<\alpha\ll\pi/2$,  for very high noise values the system forms an isotropic and homogeneous gas. Reducing $\eta$, the population undergoes two subsequent  phase transitions toward collective motion.  {\color{black}A flocking transition occurs at $\eta=\eta_{GF}(\alpha)$}: the polarization increases sharply while the global spin remains vanishingly small, Fig.~\ref{fig3alpha1}b. However, unlike all the Vicsek-like models with  periodic boundary conditions, {\color{black} the  global polarization does not align with one of the principal axis of the simulation box}. In contrast, it makes a finite angle, $\psi$,  that is set by the anticipation angle $\alpha$, Figs.~\ref{fig3alpha1}c, and ~\ref{fig3alpha1}d. 
 
 When further reducing the noise amplitude below $\eta=\eta_{FS}(\alpha)$, the flock self-organizes into a spinning phase.  {As opposed to the case where $\alpha=\pi/2$ examined in previous section,} we show in Fig.~\ref{fig3alpha1}b that $\Sigma$ increases sharply and saturates to 1 together with $\Pi$,  as $\eta$ goes to 0 deep in the spinning phase.  
 {The individual circular trajectories are now robust. {\color{black}They do not disrupt and reform intermittently; all the particles endlessly follow synchronous circular orbits, see Figs.~\ref{fig3alpha1}e, and ~\ref{fig3alpha1}f. }
 
In the spinning phase, the curvature $\kappa$ of the particle orbits has a constant value, Fig.~\ref{fig3alpha1}g. This result can be explained by considering the zero-noise limit. When $\eta=0$, a perfectly polarized state with all particles aligned is a solution of Eq.~\ref{vsalpha} for any value of $\alpha$, provided that the particles rotate at the same angular velocity $\dot \theta=\sin\alpha$.  As the particle velocity $v_0$ is a constant, the corresponding trajectories are circles of curvature $\kappa=\sin\alpha/v_0\tau$.  As seen in Fig.~\ref{fig3alpha1}h, this mean-field argument correctly accounts for our numerical results when $0<\alpha<\pi/2$. 

Figure~\ref{fig3alpha1}a shows that the phase behavior of the population is qualitatively different  when $\alpha$ approaches and exceeds $\pi/2$. The polar-liquid/flocking phase, does not exist anymore above $\alpha_{\rm T}\sim 0.9(\pi/2)$. The system undergoes a single transition from an isotropic to a spinning state. For $\alpha_{\rm T}\sim 0.9(\pi/2)$ the transition occurs through  a tricritical point, where the flocking, the spinning and the isotropic phases coexist. In addition, above $\alpha_{\rm T}$ the spinning phase displays marked qualitative differences from the one characterized above for small values of $\alpha$: 
the polarization and the spin do not saturate to 1 as $\eta\to0$ (as previously noted for $\alpha=\pi/2$). In addition the curvature of the trajectories deviates from the naive mean-field prediction, Fig.~\ref{fig3alpha1}h.

In order to elucidate this rich phase behavior, {\color{black} we now investigate the linear stability of the two ordered states. Let us first consider an ensemble of particles forming a polar liquid oriented along $\theta=0$}. Far from a transition point, the angles weakly deviate from $\theta_i=0$ and the angular dynamics reduces to:
\begin{align}
\dot\theta_i(t)\sim\sin\alpha \langle \sigma_i\rangle_{j\in\Omega_i}-\cos\alpha \langle\theta_i-\theta_j\rangle_{j\in\Omega_i}+\sqrt{2\eta}\xi_i
\label{polarphase}
\end{align}
{\color{black}In this polarized but non-spinning state, the individual spins undergo uncorrelated fluctuations}. Therefore, for any finite interaction radius, the first term on the right-hand side acts as an additional angular noise that impedes the velocity alignment, in stark contrast with the intuitive picture that one could have about the effect of anticipation. The magnitude of this effective angular noise increases with $\alpha$ which qualitatively explains why the flocking transition occurs  for smaller values of $\eta$ as $\alpha$ increases. In addition, it readily follows from Eq.~\ref{polarphase} that the effective angular stiffness, $\cos\alpha$, is negative for $\alpha>\pi/2$. Hence the homogeneous polar-liquid phase is linearly unstable to angular fluctuations.  In agreement with the phase diagram shown in Fig. 3a, no polar-liquid phase exists for $\alpha>\pi/2$. 

Let us now repeat the same analysis for the spinning phase. Introducing the angular fluctuations $\delta\theta_i\equiv\theta_i\pm t\sin\alpha$ deep in the ordered phase, the orientational dynamics reduces to:
\begin{align}
\delta\dot\theta_i(t)\sim-\cos\alpha \langle\delta\theta_i-\delta\theta_j\rangle_{j\in\Omega_i}+\sqrt{2\eta}\xi_i.
\end{align}
Again we find that the orientational fluctuations are stabilized for $\alpha<\pi/2$ and amplified otherwise. {\color{black}
This stability analysis is consistent with the strong fluctuations of the individual trajectories found for $\alpha=\pi/2$:  the spinning state is linearly  unstable for large anticipation angles. The ordered spinning phase numerically found for $\alpha \gtrsim \pi/2$ is therefore stabilized by the nonlinear nature of the spin-angle interactions and cannot be captured by a  linear-stability analysis alone.}

{\color{black} The flocking and the spinning phases are separated by another  dynamical state where the polar-liquid phase condenses into compact swarms undergoing coherent directed motion. This behavior is  characterized and illustrated  in Figs.~\ref{fig4swarm}a and ~\ref{fig4swarm}b respectively, see also Supplementary Movie~4~\cite{supp}. 
This swarming phase exists in a narrow range of the noise amplitude $\eta$, typically a few percent above $\eta_{FS}$.} Unlike the band-shape patterns observed in the standard Vicsek models, these compact swarms are {\em not} surrounded by a sea of isotropic particles, but freely propagate in an empty space.  To our knowledge, this condensation is the first evidence of the self-organisation of motile particles into compact swarms without any attractive interactions.
\begin{figure}
\begin{center}
\includegraphics[width=\columnwidth]{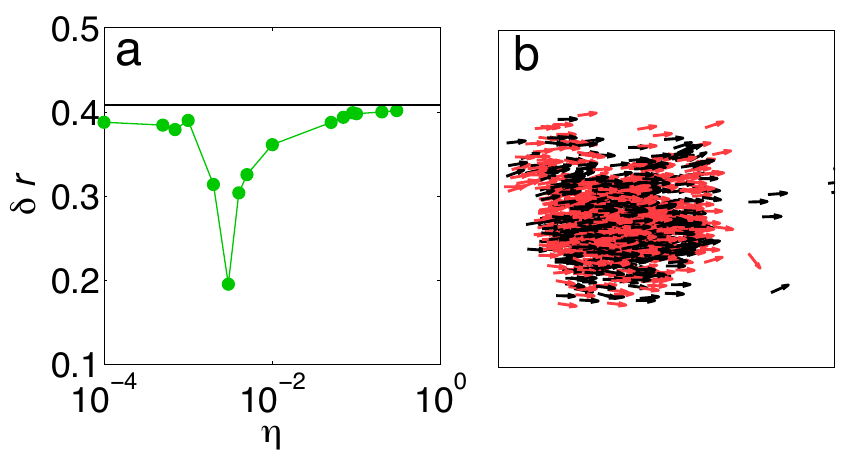}
\caption{Color online. Swarming at the onset of collective spinning. (a) The variations of $\delta r$ echo the condensation of the particles into a single compact swarm undergoing directed motion. (b) Snapshot of the instantaneous particle positions and orientations in the swarming state where $\Pi = 0.977$ and $\Sigma = 0.063$. $\alpha = 0.8\,\pi/2$, $\eta=3 \times 10^{-3}$, $\rho=8$, $v_{0}=0.5$ and $L=16$ {\color{black}($N=2048$)}. }
\label{fig4swarm}
\end{center}
\end{figure}

{\color{black}
We close this section by pointing two questions that readily arise from the above analysis. First the nature of the transition between the  four phases (gas, flocking, spinning  and swarming states) remains to be elucidated (critical, first order, or sharp cross-over). Answering this question would require studying  much larger systems and a systematic finite-size analysis~\cite{ChatePRE}, which goes beyond the scope of the present paper. A second natural question concerns the robustness of these phases with respect to the noise structure. We focused here on the case of an angular noise. Whereas the present results are expected to hold for a vectorial noise as well, the impact of more complex multiplicative noises, such as the one found to be relevant to locust swarms  in~\cite{Yates}, remains an  open question.}

\section{Conclusion and  perspectives}
We conclude this paper by addressing potential applications of our findings to multi-agent robotics. In this context, biomimetic strategies have always been  appealing to achieve collective intelligence. 
However  man-made programable units could exploit a virtually infinite number of alternative strategies to accomplish emergent tasks. In this paper,  we have demonstrated that  simple artificial interaction can be effectively used to  self-organize an entire population. By coupling the spin and the orientation of motile particles, three  symmetries can be spontaneously broken (rotation, translation and spin-reversal) thereby resulting in three  spatiotemporal patterns: homogeneous and coherent directed motion (flocking), synchronous circular motion (spinning) and condensation into a compact group propagating in a coherent fashion (swarming). 
In addition, the transition from a swarming  to a spinning state can be triggered by a minute increase of the anticipation angle $\alpha$ that characterizes the spin-orientation coupling. Therefore this sharp transition could provide a useful design rule to arrest and  disperse a motile swarm without having to stop the individual propulsion. Since  all the particles undergo synchronous circular orbits in the spinning state, a swarm of particles moving coherently  could stop and explore the neighborhood of a given spot without having to worry about possible collisions between the motile units. Experimental investigations along these lines are now needed  to address the relevance of these simple interaction rules  to devise functional robotic swarms.

\begin{acknowledgments}
DB acknowledges support from Institut Universitaire de France, and ANR project MiTra. 
\end{acknowledgments}

\end{document}